# The eth formalism in numerical relativity

R. Gómez, L. Lehner, P. Papadopoulos ‡ and J. Winicour

Department of Physics and Astronomy, University of Pittsburgh, Pittsburgh, PA 15260, USA

**Abstract.** We present a finite difference version of the eth formalism, which allows use of tensor fields in spherical coordinates in a manner which avoids polar singularities. The method employs two overlapping stereographic coordinate patches, with interpolations between the patches in the regions of overlap. It provides a new and effective computational tool for dealing with a wide variety of systems in which spherical coordinates are natural, such as the generation of radiation from an isolated source. We test the formalism with the evolution of waves in three spatial dimensions and the calculation of the curvature scalar of arbitrarily curved geometries on topologically spherical manifolds. The formalism is applied to the solution of the Robinson-Trautman equation and reveals some new features of gravitational waveforms in the nonlinear regime.

Short title: The eth formalism in numerical relativity

‡ Present address: Department of Astronomy and Astrophysics, Pennsylvania State University, University Park, PA 16802, USA

## 1. Introduction

The inspiral and merger of a binary black hole system is anticipated to be the prime source of gravitational waves for future wave detectors. Calculation of the emitted waveform, by means of emerging supercomputer technology, is the goal of the Binary Black Hole Grand Challenge [1]. In this paper we present a formalism which plays a strategic part in extracting the waveform produced in this three dimensional problem [2].

At large distances from a compact source the wavefronts of any radiation field become spherical; this leads naturally to the use of a spherical coordinate system. Indeed, spherical coordinates and spherical harmonics are standard analytic tools in the description of radiation. But the use of spherical coordinates in numerical work leads invariably to the vexing problem of coordinate singularities at the origin and along the polar axis. Finite difference approximations suffer particularly because they have no natural way of enforcing the correct boundary conditions on their solutions.

As a result, spherical coordinates have mainly been used in axisymmetric systems, where the polar singularities may be regularized by standard tricks. In the absence of symmetry, these techniques do not easily generalize and they would be especially prohibitive to develop for tensor fields.

This paper presents an approach that should be of interest to researchers working in numerical relativity in problems where spherical coordinates are a natural tool. The eth formalism [3, 4] and the associated spin-weighted spherical harmonics [4, 5] allow a simple and unified description of vector and tensor fields, without the undue complexity of traditional vector and tensor harmonics. Unfortunately, these techniques have remained obscure and unfamiliar to many researchers who would benefit from them.

We present here a finite difference version of the eth formalism, with applications to numerical relativity. The framework for the computational application of the eth formalism is developed in Sec. 2. Although our presentation is on the algorithmic level, we present enough detail to direct a well defined transition from algorithm to code development. In Sec. 3, we present tests and applications of the formalism which demonstrate both its accuracy and usefulness. The first application, the evolution of scalar waves is illustrative of the techniques necessary to carry out a 3-dimensional evolution using spherical coordinates. The second application, the numerical calculation of the curvature scalar of a topologically spherical manifold illustrates the global and tensorial features of the method. This calculation is at the heart of several important problems, such as the computation of the Hawking mass [6]. The third application, solution of the Robinson-Trautman equation, reveals new and unexpected nonlinear properties of the gravitational waveform. This work illustrates how computational relativity can greatly benefit from approaches based upon the best analytic tools.

## 2. The Eth Formalism

We present here a finite difference treatment of the sphere, based upon (i) the standard method of describing the global differentiability of functions by means of local coordinate patches, and (ii)the eth formalism for describing differentiation of fields on the sphere.

### *2.1. Patching*

Let $\theta$ and $\phi$ label the points on the sphere. Then, the real and imaginary parts of the stereographic coordinate $\zeta_N = \tan(\theta/2)e^{i\phi}$ provide a smooth coordinatization of the sphere excluding the point $\theta = \pi$ at the south pole. Similarly $\zeta_S = 1/\zeta_N$ provides a smooth coordinatization except at the north pole. These two stereographic coordinate patches are sufficient to cover the sphere. It would be possible, in principle, to cover the sphere with a number of non-singular $(\theta, \phi)$ patches, by selecting several directions $(\theta_i, \phi_i)$ on the sphere and associating with them a local coordinate patch, and then excluding from those patches the (local) polar region. Stereographic coordinates, provide the most economical choice, where only two non-singular patches are used.

A smooth scalar field $\Psi$ on the sphere may be represented in terms of smooth functions $\Psi_S$ of $\zeta_S$ on the lower hemisphere $|\zeta_S| \leq 1$ and as smooth functions $\Psi_N$ of $\zeta_N$ on the upper hemisphere $|\zeta_N| \leq 1$. The continuity of $\Psi$ is ensured by requiring $\Psi_S(\zeta_S) = \Psi_N(\zeta_N)$ at the equator but in order to ensure smoothness this equality must extend to a common overlap region about the equator. Our first task is to implement this in a finite difference treatment.

In order to construct the grid, let $(S^1, S^2)$ and $(N^1, N^2)$ be the real and imaginary parts of $\zeta_S$ and $\zeta_N$, respectively. Let the grid consist of the points $S^i = s_i/M$ and $N^i = n_i/M$, where $s_i$ and $n_i$ are each pairs of integers with $-M \leq s_i \leq M$ and $-M \leq n_i \leq M$. Thus the grid points at the intersection of the real and imaginary axes with the equator correspond to $s_i = (M, 0)$ and $s_i = (0, M)$; and $n_i = (M, 0)$ and $n_i = (0, M)$. The grid points at the poles correspond to $s_i = 0$ and $n_i = 0$. Note that our choice of rectangular grid domains is not unique [7]. Other strategies can be followed, to economize the number of points in the overlap region between the $S$ and $N$ grids, but the rectangular choice lends itself to a simple implementation at the grid boundaries.

Given a smooth function $\Psi$ on the sphere, we now give instructions for representing it on the grid. Denote the values of $\Psi_S(\zeta_S)$ at points on the $S$-grid by $\Psi_{Ss_1,s_2}$; and the values of $\Psi_N(\zeta_N)$ at points on the $N$-grid by $\Psi_{Nn_1,n_2}$. In the overlap region, the grid representation must be consistent with the condition $\Psi_S(\zeta_S) = \Psi_N(1/\zeta_S)$. However, only in exceptional cases will a point that lies on the $S$-grid also lie on the $N$-grid. To deal with this, we introduce

$$\zeta_{SNn_1,n_2} = \frac{(n_1 - in_2)M}{(n_1^2 + n_2^2)}, \tag{1}$$

which gives the value of $\zeta_S$ at a point which lies on the $N$-grid. Similarly, we define

$$\zeta_{NSs_1,s_2} = \frac{(s_1 - is_2)M}{(s_1^2 + s_2^2)}. \tag{2}$$

In the same way, in the overlap region, $\Psi_S$ and $\Psi_N$ determine the values $\Psi_{Ss_1,s_2}$ and $\Psi_{NSs_1,s_2}$ at points which lie on the $S$-grid; and $\Psi_{Nn_1,n_2}$ and $\Psi_{SNn_1,n_2}$ at points on the $N$-grid. However, the grid values $\Psi_{Ss_1,s_2}$ are weaker information than the smooth function $\Psi_S$ and do not by themselves determine the value of $\Psi$ at points on the $N$-grid. Similarly, the values $\Psi_{Nn_1,n_2}$ do not by themselves determine $\Psi$ at a point on the $S$-grid. In these cases interpolation is necessary to evaluate functions on one grid at points lying on the other grid.

In order to see how this applies to differentiation, consider the gradient of $\Psi$, with components $\partial_{S^i}\Psi_S(\zeta_S)$ in the $S$-patch; and $\partial_{N^i}\Psi_N(\zeta_N)$ in the $N$-patch. Then, at points which do not lie on the grid boundary, the partial derivatives can be approximated by centered finite differences in the standard way. For example, at the point $S^i = s_i\Delta$, the appropriate finite difference approximation is

$$\partial_{S^1}\Psi_S = \frac{\Psi_{Ss_1+1,s_2} - \Psi_{Ss_1-1,s_2}}{2\Delta} + O(\Delta^2). \tag{3}$$

For a point on the grid boundary, e.g. the point $S^1 = (1, s_2\Delta)$, we use the same approximation except that the field at the virtual grid point $s_i = (M+1, s_2)$ is approximated by the value $\Psi_{NS(M+1),s_2}$ obtained from interpolating $\Psi_N$. In order to achieve an error of second order in the derivative the interpolation error must be of third order.

This provides second order accurate finite difference representations of the gradient of $\Psi$ at all grid points and, in fact, two representations, corresponding to $\partial_{S^i}\Psi_S(\zeta_S)$ and $\partial_{N^i}\Psi_N(\zeta_N)$, at points which lie in the overlap between the two grids. The analytic expressions are related by the transformation

$$\partial_{\zeta_S}\Psi_S(\zeta_S) = \frac{\partial_{\zeta_N}}{\partial_{\zeta_S}}\partial_{\zeta_N}\Psi_N(\zeta_N). \tag{4}$$

As a check on the finite difference approximation, the finite difference equivalent of (4) must hold to second order accuracy. Equivalently, the gradient may be converted into a scalar field by taking basis components and the scalar field compared between patches. This is the strategy of the eth formalism that we shall consider later.

Second derivatives of a scalar field with respect to the coordinates may also be approximated to second order by central differencing. The interior points in each grid patch are handled by the standard techniques and the points on the grid boundary are handled by introducing a virtual grid point, as above. The following fourth order interpolation scheme ensures that both the first and second derivatives are approximated to second order accuracy at boundary points.

First, we extend the grid by one point on each coordinate direction. The points at the $S$ grid with coordinates $s_i = (M+1, s2)$ will define the computational rightmost boundary of the grid, etc.. Centered derivatives in the $S_1$ direction require the field value at the virtual grid point $s_i = (M+2, s2)$. Note that this virtual point maps into a point on the $N$ grid which has at least two neighbors on each direction who also fall inside the computational boundary of the $N$ grid, namely their grid coordinates satisfy the relations $|n_i| \leq |M+1|$.

We interpolate the value at the virtual point by the following procedure. First, we evaluate the coordinates $n_i$ of the virtual point, which according to (2) are given by $n_1 = s_1 M/(s_1^2 + s_2^2)$, $n_2 = -s_2 M/(s_1^2 + s_2^2)$. We then compute the indices $(i_b, j_b)$ of the lower left corner of the grid cell on the $N$ grid where the virtual point falls, i.e. $i_b\Delta \leq n_1 \leq (i_b+1)\Delta$ and $j_b\Delta \leq n_2 \leq (j_b+1)\Delta$. The nine grid cells whose corners are given by the points $(i,j)$ with $i_b - 1 \leq i \leq i_b + 2$ and $j_b - 1 \leq j \leq j_b + 2$ fall entirely inside the $N$ grid. We then fit a cubic Lagrange polynomial to each of the four set of points obtained by keeping the $j$ index fixed, and obtain the value of the field at the points $(s_1, j), j = j_b - 1, ..., j_b + 2$. Another cubic Lagrange polynomial is then fit through these four points to yield the value at the virtual point $(s_1, s_2)$. Note that all grid points at the boundary can be treated with the same algorithm. This method has the computational advantage that it vectorizes on machines with the appropriate hardware, such as Cray supercomputers.

We have carried out checks of the interpolation error. Even for very coarse grids, the measured convergence rate is $\approx 3.9$, in good agreement with the theoretical limit of four.

### *2.2. Eth*

The eth (ð) formalism gives a compact and efficient manner of treating vector and tensor fields on the sphere, as well as their covariant derivatives. In addition, the associated spin-weighted spherical harmonics provide a very tidy alternative to vector and tensor spherical harmonics. We give here a brief description of the basic ideas and the details of how it works in some simple cases.

The line element for the distance between neighboring points on the sphere is given by the quadratic form

$$ds^2 = q_{ab} dx^a dx^b \tag{5}$$

where the components $q_{ab}$ of the metric tensor depend upon the choice of coordinates.

In standard spherical coordinates this takes the form $ds^2 = d\theta^2 + \sin^2\theta d\phi^2$ but here we use stereographic coordinates for which

$$ds^2 = 4(1 + \zeta_S\bar{\zeta}_S)^{-2}[(dS^1)^2 + (dS^2)^2] \tag{6}$$

in the North patch, with a similar expression holding in the South patch. The eth formalism represents the metric in terms of a complex basis vector $q_a$,

$$q_{ab} = (q_a\bar{q}_b + \bar{q}_a q_b)/2, \tag{7}$$

where $q^a q_a = 0$ and $q^a\bar{q}_a = 2$. (Note that this departs from other conventions [4] to avoid unnecessary factors of $\sqrt{2}$ which would be awkward in numerical work.) Since the real and imaginary parts of $q_a$ are both vector fields of unit length, this basis cannot be assigned in a smooth manner over the entire sphere so that two patches are necessary. In the $S$ patch we make the choice $q^a_S = (1+\zeta_S\bar{\zeta}_S)(\delta^a_1 + i\delta^a_2)/2$, so that its real and imaginary parts line up with the $S$ axes. Similarly, in the $N$ patch, $q^a_N = (1 + \zeta_N\bar{\zeta}_N)(\delta^a_1 + i\delta^a_2)/2$. Any two complex bases $q_a$ and $\hat{q}_a$ are related by a unitary transformation $\hat{q}_a = e^{i\alpha}q_a$, where the phase $\alpha$ is a real valued function. In particular, in the overlap between the patches, $q^a_N = e^{i\alpha}q^a_S$, with $e^{i\alpha} = -\bar{\zeta}_S/\zeta_S$.

Introduction of this complex basis allows us to represent any vector field on the sphere, with components $v_a$, in terms of the complex scalar field $v = q^a v_a$. Here it is implicit that $v$ represents either $v_S = q^a_S v_a$ or $v_N = q^a_N v_a$, depending upon which patch is being used, with $v_N = e^{i\alpha}v_S$ in the overlap region. A field $v$ with this transformation property is called a spin-weight 1 field.

Similarly, a tensor field on the sphere, with components $w_{ab}$, can be represented by scalar fields. First we decompose $w_{ab}$ into its symmetric-trace-free part, its trace part and its antisymmetric part:

$$w_{ab} = t_{ab} + \frac{p}{2}q_{ab} + i\frac{u}{4}(q_a\bar{q}_b - \bar{q}_a q_b), \tag{8}$$

where $p = q^{cd}w_{cd}$ and $u = i(q^c\bar{q}^d - q^c\bar{q}^d)w_{cd}/2$. The (real) scalar fields $p$ and $u$ are independent of choice of basis and are called spin-weight zero fields. The symmetric, trace free tensor field $t_{ab}$ can be represented by the complex scalar field $t = t_{ab}q^a q^b$. Under change of basis we then have $t_N = e^{2i\alpha}t_S$ and $t$ is called a spin-weight 2 field. Thus an arbitrary tensor field can be represented by two real spin-weight 0 fields and a complex spin-weight 2 field. These spin-weighted fields are the irreducible representations (of the unitary group of basis transformations) contained in the tensor field $t_{ab}$. Note that $\bar{t} = t_{ab}\bar{q}^a\bar{q}^b$ alternatively represents the trace-free symmetric part of $w_{ab}$ as a field with spin-weight minus 2.

## *2.3. Derivatives of First and Second Order*

In treating derivatives of tensor fields, it is again convenient to represent them in terms of spin-weighted scalar fields. This can be accomplished by taking basis components of covariant derivatives $\nabla_a$. In the case of a scalar field, $\nabla_a \Psi = \partial_a \Psi$, which we represent by the spin-weight one quantity $\eth\Psi = q^a \partial_a \Psi$, which represents the components of the gradient in terms of a complex field. Thus, at the grid point $s_i$ in the $S$ patch, centered approximations to the derivatives give

$$\eth_S \Psi_S = \frac{1+(s_1^2+s_2^2)\Delta^2}{2\Delta}(\Psi_{S s_1+1,s_2} - \Psi_{S s_1-1,s_2} + i\Psi_{S s_1,s_2+1} - i\Psi_{S s_1,s_2-1}) + O(\Delta^2) \quad (9)$$

The overlap condition between patches is $\eth_N \Psi_N = e^{i\alpha}\eth_S \Psi_S$.

In the case of a vector field, the covariant derivative, with respect to the unit sphere metric, is

$$\nabla_a v_b = \partial_a v_b - \Gamma^c_{ab} v_c \quad (10)$$

where the connection coefficients are determined from the metric by

$$\Gamma^c_{ab} = q^{cd}(\partial_a q_{db} + \partial_b q_{ad} - \partial_d q_{ab})/2. \quad (11)$$

This is equivalent to the spin-weight 0 field $\bar\eth v = \bar q^a q^b \nabla_a v_b$ and the spin-weight 2 field $\eth v = q^a q^b \nabla_a v_b$. Here, as above, we have $v = q^a v_a$. These complex fields contain all the information about the gradient of $v_a$. Finite difference approximations can be obtained by first reexpressing

$$\bar\eth v = \bar q^a \partial_a v - \bar\Gamma v \quad (12)$$

and

$$\eth v = q^a \partial_a v + \Gamma v, \quad (13)$$

where

$$\Gamma = -q^a \bar q^b \nabla_a q_b / 2 = \zeta. \quad (14)$$

For a second order accurate finite difference approximation, the first term in either (12) or (13) may be approximated by a centered difference, as in (9), and the second term evaluated using the grid values of $\zeta$ in (14). This leads to finite difference expressions in both the $S$ and $N$ patches, satisfying $\bar\eth_N v_N = \bar\eth_S v_S$ and $\eth_N v_N = e^{2i\alpha}\eth_S v_S$ in the overlap.

To treat differentiation of two-index tensor fields, it suffices to consider trace-free symmetric tensors represented by the pure spin-weight 2 field $t = t_{ab} q^a q^b$. Then the covariant derivative of $t_{ab}$ can be expressed in terms of the spin-weight 1 field $\bar\eth t = \bar q^a q^b q^c \nabla_a t_{bc}$ and the spin-weight 3 field $\eth t = q^a q^b q^c \nabla_a t_{bc}$. These can be reexpressed as

$$\bar\eth t = \bar q^a \partial_a t - 2\bar\Gamma t \quad (15)$$

and

$$\eth t = q^a \partial_a t + 2\Gamma t,, \quad (16)$$

with $\Gamma$ given by (14). As before, these expressions have straightforward finite difference approximations in each patch, with the overlap relationships $\bar{\eth}_N t_N = e^{i\alpha}\bar{\eth}_S t_S$ and $\eth_N t_N = e^{3i\alpha}\eth_S t_S$. This procedure generalizes to treat derivatives of tensor fields with an arbitrary number of indices.

Second covariant derivatives can be treated as products of first derivatives. However, this can lead to inefficient finite difference approximations. For example, the Laplacian may be written as

$$D^2\Psi = q^{ab}\nabla_a\nabla_b\Psi = \bar{q}^a q^b\nabla_a\nabla_b\Psi = \bar{\eth}\eth\Psi. \tag{17}$$

A straightforward finite difference approximation to $\eth$ followed by an approximation to $\bar{\eth}$ would involve next nearest neighbors, whereas a central difference approximation to the second derivative would involve only nearest neighbors. Both approximations are second order accurate but the latter is clearly preferable. There are various other second derivative expressions in which the latter approach should also be invoked.

One such expression, which commonly occurs in dealing with a spin-weight 2 field $t$, is the spin-weight zero field $\bar{\eth}^2 t$. We will use this as an example of how such expressions can be approximated to second order using only nearest neighbors. In our choice of stereographic coordinates, we have

$$\bar{\eth}^2 t = \bar{q}^a\bar{q}^b\partial_a\partial_b t - 2\bar{\zeta}(\bar{\eth}t + \bar{\zeta}t). \tag{18}$$

Thus calculation of $\bar{\eth}^2 t$ reduces to the previous calculation of $\bar{\eth}t$ and the mixed derivatives $\partial_a\partial_b t$, which can all be approximated to second order by centered differences involving only nearest neighbors. At the grid boundary, this involves values at virtual grid points which can be determined to fourth order accuracy by the interpolation scheme.

Two general operators expressing arbitrary first or second order derivatives acting on any spin-weighted field are defined here. Those operators have as arguments the field values over the sphere, the spin of the field and a set of binary values that indicate the nature of the operator. These operators allow for easy and versatile numerical implementation and cover the vast majority of expressions we expect to encounter in numerical relativistic calculations.

The first order operator is defined as

$$D_1(s,\epsilon)\psi = \frac{1}{2}(1+x^2+y^2)(\partial_x\psi + i\epsilon\partial_y\psi) + s(\epsilon x + iy)\psi, \tag{19}$$

where $\zeta = x + iy$, $s$ is the spin-weight of $\psi$ and $\epsilon = 1$ gives the $\eth$ operator while $\epsilon = -1$ gives the $\bar{\eth}$ operator.

Similarly the second order operator is defined by

$$D_2(s,\epsilon_1,\epsilon_2)\psi \;=\{\frac{1}{4}(1+x^2+y^2)^2[\partial_x^2 - \epsilon_1\epsilon_2\partial_y^2 + i(\epsilon_1+\epsilon_2)\partial_x\partial_y]$$

$$
\begin{aligned}
+ \quad & \frac{1}{2}(1+x^2+y^2)[((1+\epsilon_1\epsilon_2+s(\epsilon_1+\epsilon_2))x+i(2s+\epsilon_1+\epsilon_2)y)\partial_x \\
+ \quad & ((2s+\epsilon_1+\epsilon_2)x+i(1+\epsilon_1\epsilon_2+s(\epsilon_1+\epsilon_2))y)i\epsilon_1\epsilon_2\partial_y] \\
+ \quad & s[\frac{1}{2}(\epsilon_2-\epsilon_1)+(s\epsilon_1\epsilon_2+\frac{1}{2}(\epsilon_1+\epsilon_2))(x^2-\epsilon_1\epsilon_2 y^2) \\
+ \quad & (1+\epsilon_1\epsilon_2+s(\epsilon_1+\epsilon_2))ixy]\}\,\psi. \qquad (20)
\end{aligned}
$$

The different combinations of $\epsilon_1$, $\epsilon_2$ generate the following second order operators: $D_2(1,1) \equiv \eth^2$, $D_2(1,-1) \equiv \eth\bar{\eth}$, $D_2(-1,1) \equiv \bar{\eth}\eth$, $D_2(-1,-1) \equiv \bar{\eth}^2$.

## 3. Applications and Tests

### *3.1. Wave Evolution*

An important first application and test of the approximation scheme developed in the previous section is the numerical solution of the 3-D wave equation in spherical coordinates. We start with the the discretization of the Laplace operator on the sphere. In terms of the complex coordinate $\zeta$, we have

$$
D^2\Psi = D_2(0,1,-1)\Psi = (1+\zeta\bar{\zeta})^2\partial_\zeta\partial_{\bar{\zeta}}\Psi. \qquad (21)
$$

Since $D^2$ is a scalar operator, (21) gives the same value in the overlap region when evaluated in either the $S$ or $N$ coordinates.

In retarded time coordinates the wave equation $\Box\,\Phi = 0$ takes the form

$$
2g_{,ur} - g_{,rr} + \frac{D^2 g}{r^2} = 0, \qquad (22)
$$

where $g = r\Phi$. We solve (22) by an explicit marching algorithm, from $r = 0$ to null infinity [8].

The convergence and stability of the marching algorithm is dictated by the Courant-Friedrichs-Lewy (CFL) condition (which requires that the numerical domain of dependence includes the physical domain of dependence). In the present case, this requirement is strongest at the vertex of the outgoing cones. It limits the timestep by

$$
\Delta u \leq k\Delta r\Delta s_1\Delta s_2, \qquad (23)
$$

where $k$ is a number of order one, related to the details of the startup procedure near the vertex $r = 0$. To good accuracy, our numerical investigations give the value $k = 1$ for stable evolution.

A good test of the patching scheme focuses on the accuracy of the numerical evolution. The linearity of the problem provides a useful family of exact solutions

$$
G_{lm}(u,r,\zeta,\bar{\zeta}) = \frac{r^{l+1}}{(u+1)^{l+1}(u+2r+1)^{l+1}} Y_{lm}(\zeta,\bar{\zeta}). \qquad (24)
$$

Expressions for the standard and spin-weighted spherical harmonics in stereographic coordinates can be found in [9, 4]. The characteristic scheme evolves a global solution to the wave equation and thus it is possible to check the global numerical error. The $l_2$ norm of the error, for the evolution of initial data given by an $l = 2, m = 2$ harmonic, converges to zero with a measured 2.02 convergence rate.

Figure 1 shows a sequence of snapshots of the radiation patterns, as seen at null infinity, for initial data which consist of a localized pulse in the North patch traveling towards the origin. Notice the smooth propagation of the field across the coordinate patches as the evolution progresses.

### *3.2. Calculation of Scalar Curvature*

As a tensorial illustration of the forgoing methods, we consider a problem which arises in many different contexts in general relativity: Given the metric $h_{ab}$ of a topological sphere, calculate the scalar curvature. There are many ways to formulate this problem analytically. Here we present a treatment, in terms of an auxiliary unit sphere metric, which is flexible enough to be of broad usefulness in numerical applications.

The metric is uniquely determined by its unit sphere dyad components $K = h_{ab}q^a\bar{q}^b/2$ and $J = h_{ab}q^aq^b/2$, which are of spin weight zero and two respectively. The dyad components of the inverse metric are $h^{ab}q_a\bar{q}_b = 2K/H$ and $h^{ab}q_aq_b = -2J/H$, where $H = (K^2 - J\bar{J}) = det(h_{ab})/det(q_{ab})$. Also note that in our conventions the alternating tensor of the unit 2-sphere is $\epsilon_{ab} = iq_{[a}\bar{q}_{b]}$ and $(\bar{\eth}\eth - \eth\bar{\eth})\eta = 2s\eta$ for a spin-weight $s$ field $\eta$.

Let $D_a$ represent the covariant derivative associated with $h_{ab}$ [10].Then its curvature determines the commutator

$$(D_aD_b - D_bD_a)q_{cd} = h^{ef}(R_{abcf}q_{ed} + R_{abdf}q_{ce}), \tag{25}$$

where, in two dimensions, $R_{abcd} = Rh_{a[c}h_{d]b}$. We let the tensor field

$$C^c_{ab} = \frac{1}{2}h^{cd}(\nabla_a h_{db} + \nabla_b h_{ad} - \nabla_d h_{ab}) \tag{26}$$

represent the difference between the connection associated with $D_a$ and the unit sphere connection, e.g. $(D_a - \nabla_a)v_b = -C^c_{ab}v_c$. Then by contracting (25) with $q^a\bar{q}^bq^cq^d$, we obtain an expression for the curvature scalar in terms of the dyad components of $C^c_{ab}$,

$$RJ = \bar{\eth}C - \eth B + \frac{1}{2}(C\bar{A} + AB - B^2 - \bar{B}C), \tag{27}$$

where $A = q^aq^b\bar{q}_cC^c_{ab}$, $B = q^a\bar{q}^bq_cC^c_{ab}$ and $C = q^aq^bq_cC^c_{ab}$.

Next, we use (26) to relate these to the metric components,

$$A = \frac{1}{H}(2K\eth K - K\bar{\eth}J - \bar{J}\eth J)$$

$$B = \frac{1}{H}(K\bar{\eth}J - J\eth\bar{J})$$
$$C = \frac{1}{H}(K\eth J - 2J\eth K + J\bar{\eth}J). \qquad (28)$$

Substitution into (27) then yields the expression (29) for the curvature scalar, after canceling a common factor of $J$ from both sides:

$$R = \frac{1}{2H}\{2K - \eth\bar{\eth}K + \bar{\eth}^2 J + \frac{1}{2H}[2(\eth K)(K\bar{\eth}K - K\eth\bar{J} - J\bar{\eth}\bar{J})$$
$$+ (\bar{\eth}J)(\bar{J}\eth\bar{J} + \frac{K}{2}\bar{\eth}\bar{J}) + (\eth\bar{J})(J\bar{\eth}\bar{J} - \frac{K}{2}\bar{\eth}J)]\}$$
$$+ c.c. \qquad (29)$$

where $H = (K^2 - J\bar{J})$. The derivation of (29) is quite laborious and it would be extremely useful to develop symbolic techniques to carry out general dyad decompositions and reduce covariant derivatives to eth operations. In the form (29), the curvature scalar may be computed to second order accuracy using the finite difference techniques presented above.

In the particular case where $det(h_{ab}) = det(q_{ab})$, as in the Bondi [11] formalism, $H = 1$ and the metric is uniquely determined by $J$. In this choice of gauge the curvature formula simplifies considerably,

$$R = 2K - \eth\bar{\eth}K + \frac{1}{2}\left(\bar{\eth}^2 J + \eth^2 \bar{J}\right) + \frac{1}{4K}\left(\bar{\eth}\bar{J}\eth J - \bar{\eth}J\eth\bar{J}\right). \qquad (30)$$

Note that in (30) the complex conjugation has been explicitly carried out.

As a test, we choose $J$ to be a simple spin-weight 2 harmonic, e.g., $J = c\,\eth^2 Y_{20}$, which facilitates the calculation of exact expressions for the curvature. Scaling the metric data with the parameter $c$ generates arbitrarily non linear curvature expressions which are used to check the numerical implementation of (30).

An interesting check of the numerical procedure utilizes the global geometrical properties of curvature on spherical surfaces. The Gauss-Bonnet theorem for spherical topology implies

$$I = \oint R dS = 8\pi. \qquad (31)$$

A global test of the computation of the scalar curvature can thus be performed by integrating the numerically evaluated curvature over the entire sphere. The surface area element is $dS = 2i\,H^{1/2}(1 + \zeta\bar{\zeta})^{-2} d\zeta d\bar{\zeta}$. The integration must take into account the overlap between coordinate patches. A simple and natural choice is to use the equator $\zeta\bar{\zeta} = 1$ as the smooth and symmetric boundary of the integration within each patch. A second order integration scheme in two dimensions is the simple four point center average method. We must pay careful attention to the boundary cells, i.e. those which overlap the equator. Their number scales like $1/\Delta$ and if we were to omit their contribution to the integral the order of accuracy would be reduced by one. To take them into account, the interior area of each boundary cell is approximated assuming that the equator looks like a straight grid line locally. The centered four point average of the cell is then weighted by its interior area divided by $\Delta^2$.

Figure 2 presents the Gauss-Bonnet global test and confirms the convergence rate of 2.

### *3.3. Robinson-Trautman Solutions*

Robinson-Trautman spacetimes [12] contain purely outgoing gravitational radiation, which decays to leave a Schwarzschild-like horizon [13, 14]. The Bondi news function for these spacetimes is given by [15]

$$N = \frac{1}{2}\, \mathcal{W}^{-1}\eth^2\mathcal{W}, \tag{32}$$

where $\mathcal{W}(u,\zeta,\bar\zeta)$ satisfies the Robinson-Trautman equation

$$12\partial_u\mathcal{W} = \mathcal{W}^3(\eth^2\mathcal{W}\bar\eth^2\mathcal{W} - \mathcal{W}\eth^2\bar\eth^2\mathcal{W}). \tag{33}$$

The real and imaginary parts of the news function represent the two independent polarization modes of the radiation. The Bondi mass is given by the solid angle integral $M = (1/4\pi)\oint \mathcal{W}^{-3}d\Omega$. The Schwarzschild spacetime (for a unit mass black hole) corresponds to the solution $\mathcal{W} = 1$. The linearized perturbation $\mathcal{W} = 1 + A(u)Y_{\ell m}$ decays exponentially according to $A = A(0)e^{-u\,\ell(\ell+1)(\ell^2+\ell-2)/12}$.

Accurate numerical evolution of a fourth order parabolic equation, such as (33), by means of an explicit finite difference scheme is a challenge because the $CFL$ condition requires that the time step $\Delta u$ scale as the fourth power of the spatial grid size. Nevertheless, we have applied the $\eth$ formalism to obtain an efficient, second-order accurate evolution algorithm, based upon a three time level Adams-Bashford [16] scheme with predictor $(\tilde{\mathcal{W}})$ given by

$$\tilde{\mathcal{W}}(u+\Delta u) = \mathcal{W}(u) + \frac{\Delta u}{2}\partial_u[3\mathcal{W}(u) - \mathcal{W}(u-\Delta u)], \tag{34}$$

and corrector

$$\mathcal{W}(u+\Delta u) = \mathcal{W}(u) + \frac{\Delta u}{2}\partial_u[\mathcal{W}(u) + \tilde{\mathcal{W}}(u+\Delta u)] + O(\Delta u^3). \tag{35}$$

Here the $\partial_u$ terms are to be considered reexpressed in terms of $\eth$ operators using (33). The second order convergence of numerical solutions was confirmed in the perturbative regime using solutions of the linearized equation.

This algorithm reveals some new and unanticipated features of the gravitational waveforms in the nonlinear regime. Figures 3 and 4 illustrate the Bondi news function for the Robinson-Trautman space-time with initial data in the odd parity $\ell = m = 3$ mode

$$\mathcal{W}|_{u=0} = 1 + \lambda \Re[Y_{33}], \tag{36}$$

for the case $\lambda = 0.89$. In order to provide a physically intrinsic time dependence, the news function is plotted versus Bondi time $u_B$, which is related to the coordinate time $u$ by $du_B/du = \mathcal{W}$.

Figure 3 shows surface plots at representative times of the nonlinear contribution to the news function, *i.e.* $\Delta N = N(\lambda) - \partial_\lambda N|_{\lambda=0}\lambda$. (The plots show the real part of $\Delta N$ in the south patch). The smoothness of the plots indicate the effectiveness of the algorithm in handling the high spherical harmonics generated by the nonlinearity. The highest harmonics damp quickly and the system asymptotically approaches a monopole-dipole combination at late times.

Figure 4 displays the time evolution of the news function at a representative point on the equator. The graphs reveal an oscillatory behaviour qualitatively quite different from the pure exponential decay of the linearized solution. In addition, the behavior of the two polarization modes is quite different. Approximately 5% of the initial mass is radiated away during this simulation, which ends at the time of black hole formation. This is in the range of energy expected to be radiated during the inspiral of a binary black hole system. In this regime, we have found (by varying the value of $\lambda$) that these oscillatory waveforms cannot be well approximated even by second order perturbation theory, which emphasizes the importance of an accurate computational treatment.

## Acknowledgments

We are thankful for research support under NSF Grants PHY95-10895 and PHY93-18152/ASC93-18152 (ARPA supplemented) and for computer time made available by the Pittsburgh Supercomputing Center.

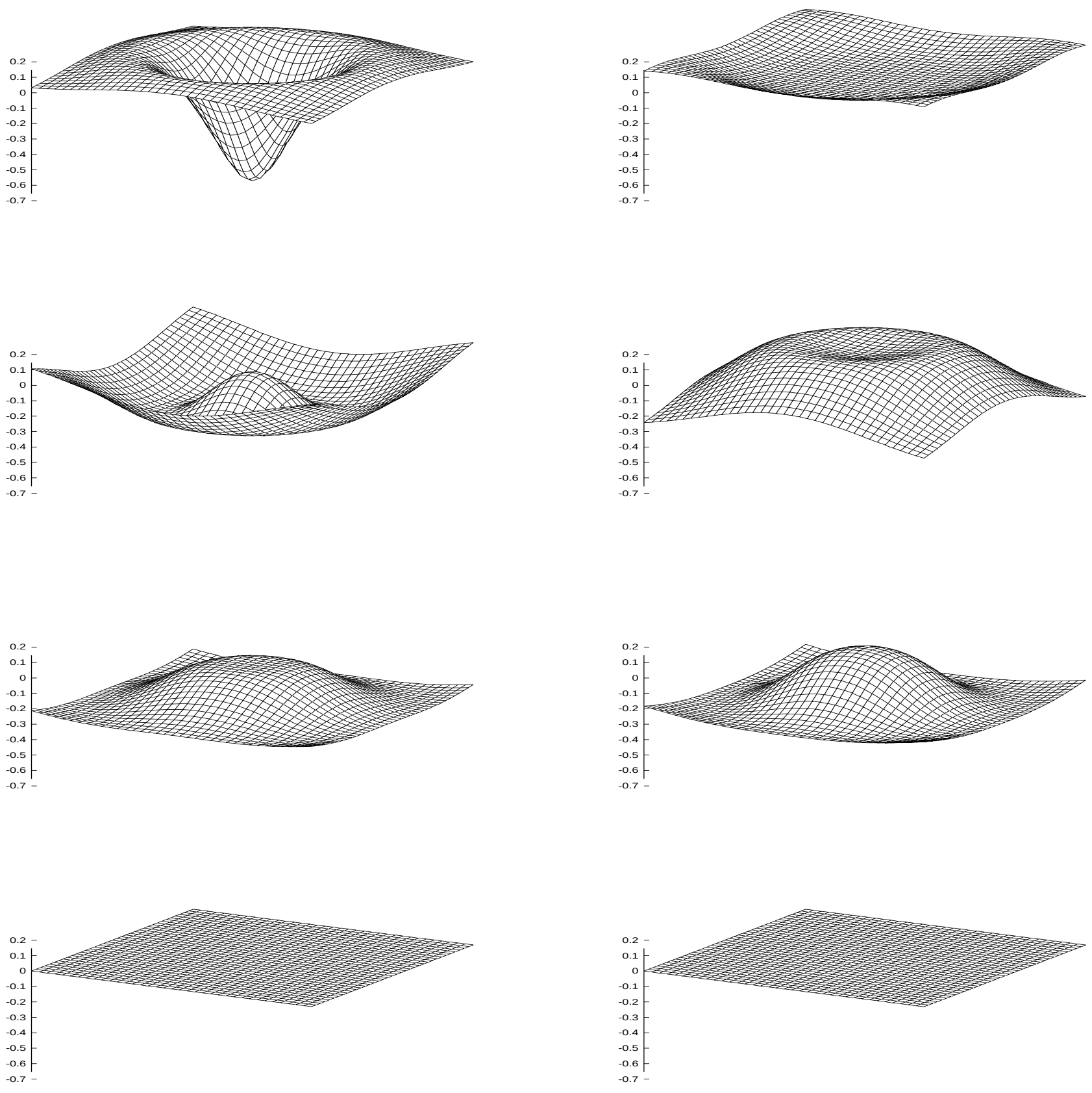

**Figure 1.** A sequence of radiation patterns at infinity for initial data consisting of a localized pulse which is traveling towards the origin. The left column represents the north patch, the right column the south patch. The evolution proceeds from top to bottom. The first pair is at retarded time $u = 0.04$ and displays the initial radiation resulting from scattering of the pulse off the centrifugal barrier created by its high angular momentum ($l$ multipole values). This radiation shows up in the same hemisphere as the initial data. At the next two retarded times, $u = 0.16$ and $u = 0.32$, radiation appears in all directions as the pulse crosses the origin defining retarded time. Finally, at $u = 0.8$ the pulse has completely passed by the origin and the radiation ceases. The grid size is $M = 18$, i.e. 37x37 points per patch.

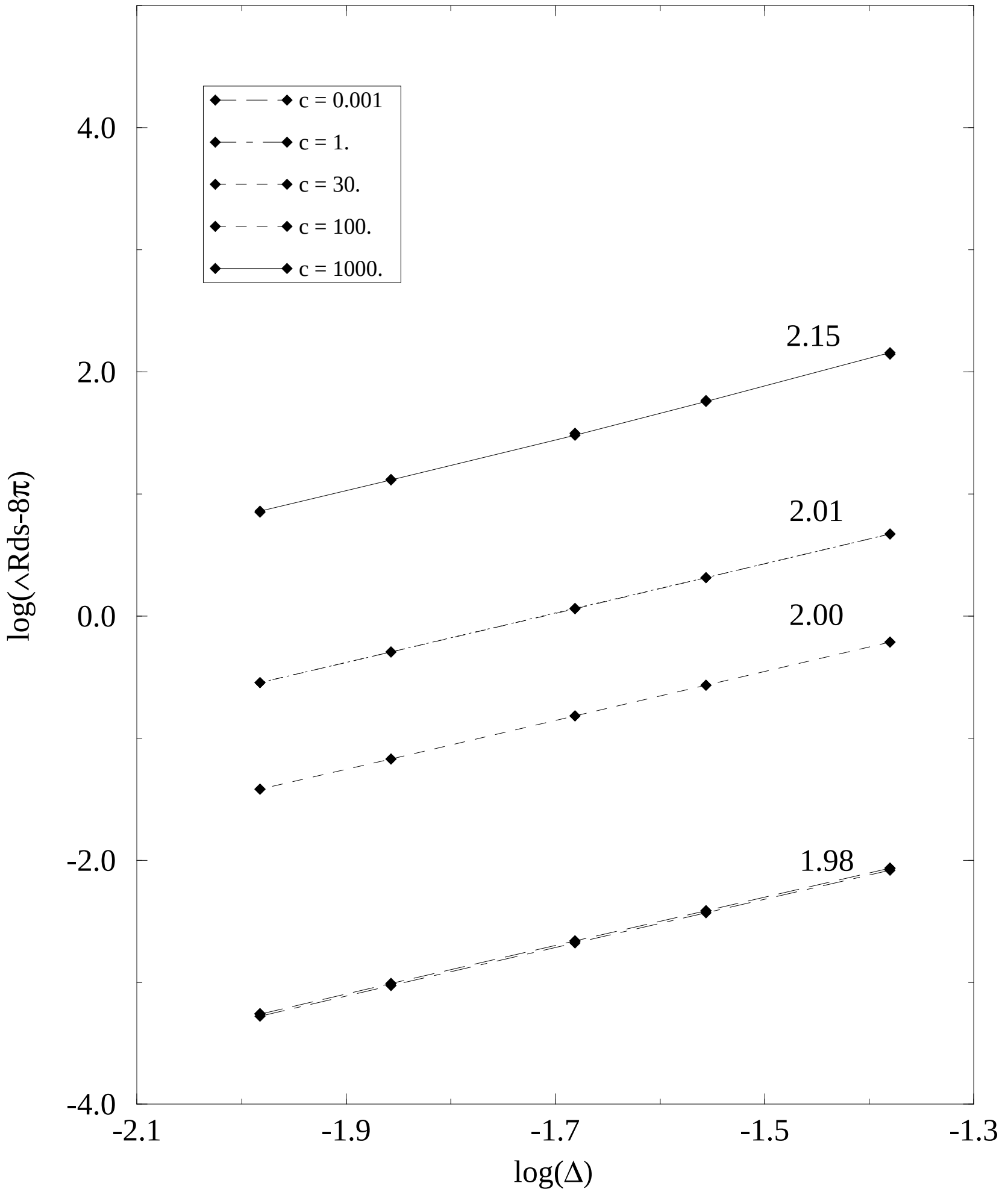

**Figure 2.** The global Gauss-Bonnet convergence test. The figure displays the convergence of the numerically evaluated integral $I = \oint R ds$ to the exact value of $8\pi$. The seed for the metric data is an $l = 2$, $m = 0$ harmonic and it is scaled over six orders of magnitude. The convergence is uniformly second order. Note that for very low amplitudes the error is dominated by the surface integration procedure and is independent of the data.

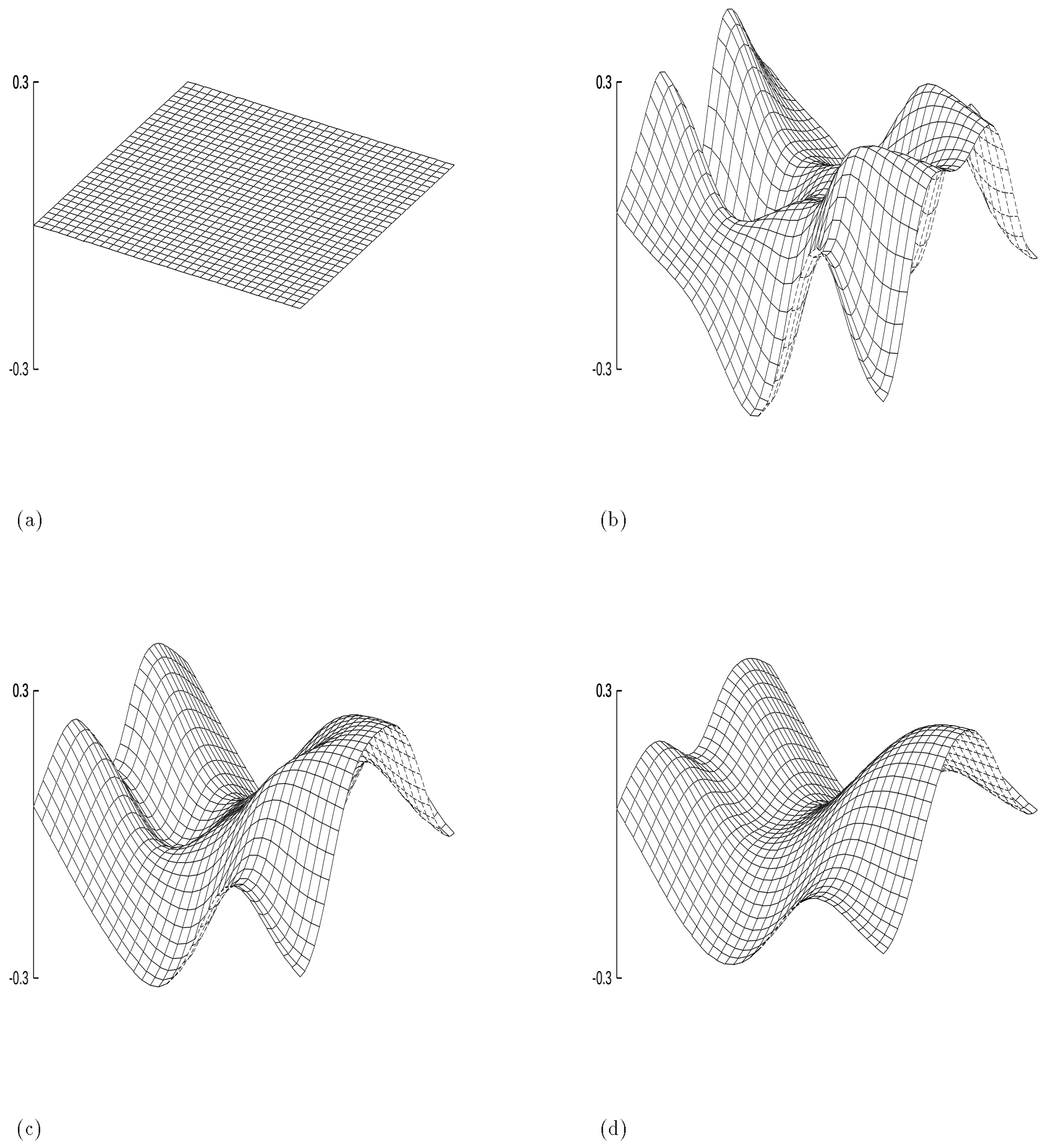

**Figure 3.** Comparison of linear versus non-linear evolution of outgoing radiation fields. The graphs show the difference between the Bondi News for the perturbative and the numerically evolved spacetimes for (a) $u_B = 0.$, (b) $u_B = 0.07$, (c) $u_B = 0.14$ and (d) $u_B = 0.25$. At early times ($u_B = 0.07$), a complex angular structure reveals the presence of high spherical harmonics which damp in a short period of time. At later times ($u_B = 0.25$), the system approaches a monopole-dipole combination.

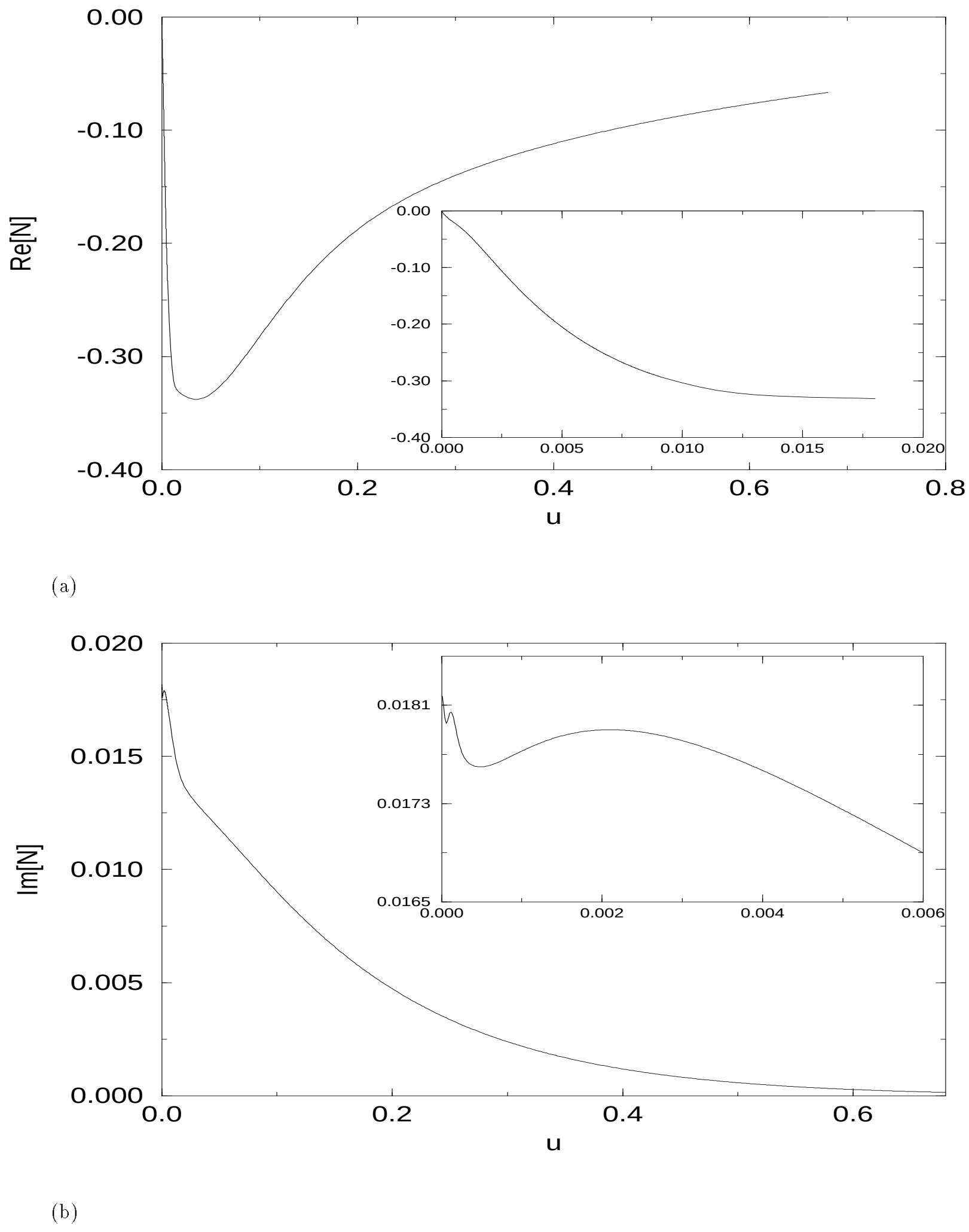

**Figure 4.** The Bondi News function at the point $(\theta, \phi) = (\pi/2, \pi/2)$, for initial data corresponding to $Y_{33}$. Both the real (a) and imaginary part (b) are shown. A perturbative calculation to first order predicts that the real part at this point is zero, with the imaginary part decaying exponentially. In contrast, the real part of the non-perturbative News is markedly different from zero, and both components show oscillations at early times.